# PREPRATION OF SILVER NANOPARTICLES DOPED PVDF: FORMATION OF PIEZOELECTRIC POLYMORPH

Dipankar Mandal[1,2], Sun Yoon[1], Kap Jin Kim[1],*

[1] Department of Advance Materials Engineering for Information and Electronics, College of Engineering, Kyung Hee University, 1 Seocheon-dong, Giheung-gu, Yongin-si, Gyeonggi-do 446-701, Republic of Korea
[2] Department of Physics, Jadavpur University, Kolkata 700032, India
* Corresponding author (kjkim@khu.ac.kr)

**Keywords**: *PVDF, β-phase, Silver nanoparticle*

**Abstract**

The preparation of polymorphism control of Poly(vinylidene fluoride) (PVDF) by silver nanoparticles (Ag NPs) is investigated. The Ag NPs were prepared by simple one step process from $AgNO_3$, where N,N-dimethylformamide (DMF) act as reducing agent as well the solvent of the host polymer, PVDF. It was observed that PVDF is one of the best stabilizers of Ag NPs. The thick films (10-20 μm) were prepared by simple solution casting followed by solvent drying and crystallizing PVDF polymorph. Here we observed that PVDF polymorph can be control by the content of the Ag NPs regardless of other processing conditions. The formation piezoelectric polymorph (β phase) by adequate amount of Ag NPs doping in PVDF was explained by the specific interaction between the surface charge of the Ag NPs and electric dipoles ($CF_2$ dipoles) comprising in PVDF. In this work we also address the suitable technique for correct crystallographic phase identification in PVDF, as a large number of works in this field have already been misled. The significant higher temperature shift of melting temperature of β-phase was observed by Ag NPs doping, which has prime importance in diverse fields of electronic applications, *i.e.* IR-sensors, piezoelectric and pyroelectric sensor, transducers as well as actuators.

**1 Introduction**

The Poly(vinylidene fluoride) (PVDF) has been extensively studied semicrystalline polymer since the finding of its piezoelectricity [1]. It has diverse crystalline phases (α, β, γ, δ, and ε) depending on the crystallization conditions. The α phase ($TGTG$) is non polar. The β phase has all-trans ($TTTT$) planner zigzag structure and the dipole moments of the two C-F and C-H bonds add up in such a way that the monomer get an effective dipole moment in the direction perpendicular to the carbon backbone. Therefore, the β phase has the largest spontaneous polarization per unit cell and thus exhibits the most superior ferroelectric and piezoelectric properties. The γ and δ phases are polar but their dipole moment is significantly smaller [2]. Therefore, PVDF containing β phase is a prime interest for electronic applications, i.e. piezoelectric or pyroelectric sensors, microwave transducers as well as non-volatile memories. The as-cast PVDF film is known to predominantly exhibit the α phase. Over the decades, the research has been focused to induce the β phase in PVDF by several methods like mechanical stretching [3], application of high pressure [4], melt-quenching [5], poling under high electric field and tension [6], electrospinning [7, 8], with incorporation of TrFE units [9], and so on.

In this work, the electroactive β phase is induced by silver nanoparticles (Ag NPs) doping in PVDF, The Ag NP doped PVDF thick films (t~10-20 μm) were prepared by a solution casting techniques for different Ag NP concentrations and the associated properties were investigated by several means. We also address the suitable technique for correct phase identification in PVDF, as a large number of works in this field have already been misled. The significant higher temperature shift of melting temperature of β-phase was observed by Ag NPs doping, which has prime importance in diverse fields of electronic applications, *i.e.* IR-sensors, piezoelectric and pyroelectric sensor, transducers as well as actuators.

## 2 Experimental Details

The Ag NPs were by one step procedure by dissolving solid $AgNO_3$ in 6 wt% (w/v) PVDF-DMF solution to make the concentrations: 2.3mM (sample: Ag2.3), 4.7 mM (sample: Ag4.7) and 7.0 mM (sample: Ag7.0). Different $AgNO_3$ were chosen to get optimum desirable properties of Ag NPs doped PVDF films. The solutions are stirred at 60°C for one day and then drop casted on clean glass slides followed by vacuum drying at 120°C one week. Afterwords, the thick films are lifted up from the glass slides. Morphology and NPs dispersion was investigated by Field Emission Scanning Electron Microscopy (FE-SEM) (SUPRA$^{TM}$ 50/50 VP), operated by an acceleration voltage of 10 kV. The Fourier Transform Infrared (FT-IR) spectroscopy was collected using a Bruker IFS 66v/S spectrometer.

## 3 Results and Discussion

In prior to solution casting, Ag NPs formation was confirmed by UV-Vis spectroscopy by detecting surface Plasmon peak around 405 nm and it was also confirmed for thick film samples. The FE-SEM image of the sample Ag7.0 is shown in fig.1. The monodispersion of Ag NPs were detected, no aggregation was found. Well grown crystalline lamer were also visible. From the X-ray Photoelectron Spectroscopy (XPS), the metalltic (Ag°) was confirmed, indicates that DMF act as reducing agent. The crystalline polymorph was investigated by FT-IR spectroscopy, shown in fig. 2 (a). The main characteristic vibrational band of β phase at 1275 cm$^{-1}$ is become prominent for Ag NPs doped PVDF samples, which is absent in Net PVDF. It seems that very small amount of $AgNO_3$ (2.3 mM in PVDF-DMF solution) can induce β phase, however non polar α phase are still present. All characteristic bands of α phase are completely diminishes for sample Ag 4.7 (4.7 mM $AgNO_3$ in PVDF-DMF solution) and Ag 7.0 (7.0 mM $AgNO_3$ in PVDF-DMF solution).

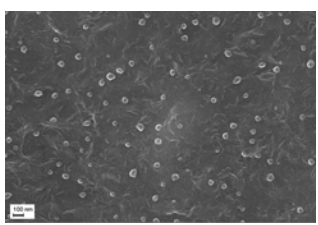

Fig. 1. FE-SEM image (magnification:100 kX) of Ag7.0.

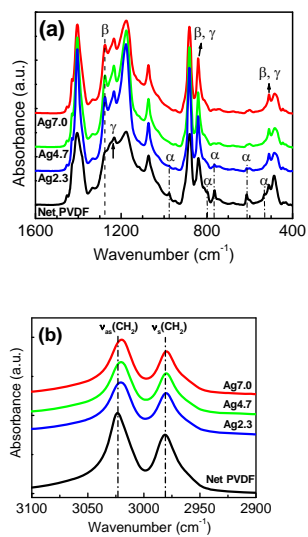

Fig. 2. FT-IR spectra of Net PVDF, Ag2.3, Ag4.7, Ag7.0 in different frequency regions; (a) 1600 to 400 $cm^{-1}$; (b) 3100 to 2900 $cm^{-1}$.

## 4 Conclusions

Our results indicates that by desirable amount of Ag NPs doping in PVDF can give rise the electroactive polar β phase due to interactions of Ag NPs and $CF_2$ dipoles existing PVDF. This might be due to the electronegativity of the fluorine and surface charge distribution of the Ag NPs , some specific interaction takes place which eventually give rise all trans confirmation in PVDF.

The asymmetric and symmetric stretching of the $CH_2$ vibrational bands (fig. 2b) towards lower energy is one of the suitable indications of the Ag NPs with PVDF.

Our final goal is to fabricate the energy harvesting devices based on Ag NPs doped PVDF films.

**Acknowledgement:** This work was supported by the NRF(R11-2005-065) and the MKE(Grant No.10033449).